# Synthesis of metallic nanoparticles for heterogeneous catalysis: Application to the Direct Borohydride Fuel Cell

Fabrice Asonkeng[1,2*], Gaël Maranzana[2], Julien Proust[1], Manuel François[3], Léa Le Joncour[3], Jérôme Dillet[2], Sophie Didierjean[2], Guillaume Braesh[4], Marian Chatenet[4], Thomas Maurer[1*].

[1] Laboratoire Lumière, nanomatériaux & nanotechnologies – L2n, Université de Technologie de Troyes & CNRS ERL 7004, 12 rue Marie Curie, 10000 Troyes, France

[2] Université de Lorraine, CNRS, LEMTA, UMR 7563, 54504 Vandoeuvre Les Nancy, France

[3] LASMIS, Université de Technologie de Troyes, 12 rue Marie Curie, 10000 Troyes, France

[4] Univ. Grenoble Alpes, Univ, Savoie Mont Blanc, CNRS, Grenoble INP (Institute of Engineering and Management Univ. Grenoble Alpes). LEPMI, 38000 Grenoble, France

**ABSTRACT:** Until now, the fabrication of electrocatalysts to guarantee long life of fuel cells and low consumption of noble metals remains a major challenge. The electrocatalysts based on metals or metal oxides which are used today are limited by the complexity of their synthesis processes and require several steps before depositing the catalysts on the substrate. Herein is described a chemical synthesis process that consists of a single-step synthesis and direct deposition of catalysts nanoparticles such as gold (Au), palladium (Pd) and platinum (Pt) in the thickness of a carbon-fibers-based porous transport layer (PTL). The synthesis process essentially consists of dissolving in the same PGMEA (Propylene glycol methyl ether acetate) solvent a metal precursor ($HAuCl_4$ or $PdNO_2$ or $PtCl_4$) and a homopolymer PMMA (Polymethylmetacrylate), then the metal solution is deposited on the surface of the PTL after cleaning. Special emphasis is made on Pt-based materials. The obtained PTL-supported nanoparticles were firstly characterized by scanning electron microscopy (SEM) to evaluate their morphology, and then X-Ray diffraction (XRD) to observe the crystal phases. To validate the methodology, Pt-coated PTL materials have been used as anode for the borohydride oxidation reaction (BOR) in a direct borohydride fuel cell (DBFC) and compared to a state-of-the-art nickel electrode. There is an optimum loading of platinum (below 0.16 mg Pt/$cm^2$) which constitutes the best compromise between power density and faradic efficiency for the borohydride oxidation reaction (BOR). Thanks to this low Pt loading, hydrogen evolved during the anodic reaction is completely valorized. These electrodes combine the advantages of high-performance with a very low metal loading, hence lowering materials'cost.

## 1) INTRODUCTION

Fuel cells are ranked among the key technologies in 2020 for emerging applications in portable and mobile electronics, where there is an increasing demand for high energy density power supplies. As such, intensive research has been carried out and is still ongoing to overcome technological barriers and commercialize this carbon-neutral technology. Among the different types of fuel cells developed to date, polymer electrolyte fuel cells (PEFCs) exhibit the best performance in terms of power density at relatively low operating temperatures (< 80°C) **[1].**

PEFCs using hydrogen as a fuel have experienced great progress in the last decades, but their commercialization remains limited so far, notably because of the problem of hydrogen storage as well as related safety concerns **[2, 3].** Hydrogen must be stored in tanks in compressed or liquefied form (which involves safety issues) or, in hydrogen storage alloys **[4].** None of these hydrogen storage methods is suitable for portable applications, due to their low volumetric and gravimetric efficiency **[5-9].** In order to overcome these difficulties, the gaseous fuel has been replaced by liquid methanol to feed so-called Direct Methanol Fuel Cells (DMFCs). This technology has attracted research attention for portable applications because of their easy fuel handling and improved storage safety compared to hydrogen **[10, 11].** However, several challenges hinder their commercialization, such as poor cell performance due to their low efficiency **[12]** and their need for high loadings in Pt-group metal (PGM) catalyst, leading to inacceptable electrode cost **[13, 14].**

Direct borohydride fuel cells (DBFCs) are recognized as an attractive and promising technology for portable and mobile applications, owing to their high theoretical voltage (1.64 V using $O_2$ as oxidant) and high energy density **[19, 20].** In addition, their strongly-alkaline anode compartment ($NaBH_4$ must be stabilized in high pH anolyte solutions **[44, 45]**) allows the use of non-precious materials (such as Ni and Cu) as catalysts for the borohydride oxidation reaction (BOR) **[21, 22, 46].** However, the electrodes produced until recently are not really optimized in terms of catalyst material, loading and shaping **[23]** even if they could sometimes show promising results with the production of anodic current at low potentials values **[24, 25]**, an endeavor for high-efficiency DBFCs.

Recently, Oshchepkov et al. have obtained performances surpassing the catalysts based on Pt in these low potential regions, by electrodepositing Ni nanoparticles on a high surface area carbon support **[26].** These results were obtained with good control of the surface finish of the Ni particles, and notably on the extent of surface oxidation of the electrodeposited nickel catalyst. This state of surface had not been taken into account in previous works, resulting in poor results with Ni-based catalysts **[47, 48].** Strongly-oxidized Ni is indeed inadequate for the borohydride oxidation reaction (BOR) in terms of initiation potential and faradic efficiency in the low potential region ($E$ < 0 V *vs* RHE) and a metallic state is required to reach promising performance. Braesch et al. have further improved the performances of such Ni-based anodes by electro-depositing these metallic nickel nanoparticles on 3D Ni structures **[27].** Using an all-metal support provides increased electronic conductivity over carbon supports, and the entire support can be active towards the borohydride oxidation reaction (BOR). However, using these Ni-based anodes results in a low faradic efficiency of 50%, due to their inability to valorize the hydrogen produced during the BOR **[26, 27, 28].**

The demands for suitable electrocatalysts to achieve the best performance in DBFC are still pending. To date, most of the reports have widely used chemical synthesis techniques such as polyol **[29]**, microemulsion **[30]**, impregnation **[31]**, colloidal methods **[32, 33]**, to prepare catalysts on a carbon support, mainly because of its conductive nature to allow good control of morphology and size **[34].** However, these techniques require the presence of surfactant species to achieve the desired size and morphology of catalysts; removing these surfactants is tedious and time-consuming **[35]**, and can sometime destabilize / reorganize the nanoparticles, which is of course not desired. Leaving the surfactant to preserve the morphology and definition of the nanoparticles is not an option, as surfactant species block most of the active sites of the catalyst, thus reducing the electrocatalytic performance **[36]**. Another technique for preparing a catalyst which is more widely used in the literature is electrochemical deposition, as introduced above for nickel-based electrodes. Electrochemical deposition can be a more advantageous technique compared to chemical synthesis technique, because it enables to deposit the catalysts directly on a conductive support at places having both ionic and electronic accessibility, which ensure very high utilization factor of the catalyst, **[37, 38, 49, 50]**. Another advantage of electrodeposition is the ability to easily adjust the characteristics of catalysts by simply changing the deposition parameters **[39]**. Unlike the chemical synthesis technique, which requires several steps for the preparation of catalysts, electrochemical deposition is a rapid procedure. However, this technique is still expensive and may be awkward to prepare large size electrodes that could be suited in industrial fuel cells or electrolyzers.

Thus, recent studies tend to indicate that the deposition of catalysts on the surface and in the volume of porous transport layers (PTL, typically gas diffusing layers, GDL), is complex and requires several steps. The colloidal synthesis technique used so far consists of synthesizing the particles first, before depositing them on the support material. Recently, our group proposed innovative routes to synthesize organized Au nanoparticles taking advantage from block copolymer properties and microsegregation **[51, 52, 53, 54]**. In continuity with these studies a simple and rapid chemical synthesis process is proposed, for which the synthesis and direct deposition of the catalysts on a support material is performed in a single step (Figure 1 – illustrated for Au nanotriangles). These gold nanotriangle catalysts were deposited in the thickness of the carbon PTL in a first attempt; then, Pd and Pt nanoparticle catalysts were prepared following the same process. The nanocatalysts are created in real time during deposition and dispersed uniformly on the surface of the fibers and in the volume of the porous carbon substrate, which is the originality of this process. Special emphasis is made on Pt-based materials hereafter. The obtained materials were firstly characterized by scanning electron microscopy (SEM) to evaluate their morphology, and then X-Ray diffraction (XRD), to observe the crystal phases of the platinum nanoparticles. The Pt catalysts that were deposited in the thickness of the carbon support were used as an anode for the borohydride oxidation reaction (BOR) in DBFCs. Three loadings (0.16; 0.32 and 0.48 mgPt/cm$^2$) were used to compare faradic efficiency and cell performance to that of the nickel electrode tested in Refs **[26-28]**.

**2) EXPERIMENTAL**

### 2.1) Materials and methods

**➔ Cleanning the carbon substrate (PTL)**

The chosen carbon PTL, (Sigracet – SGL – 25 AA) with a thickness of 260 μm, does not bear any microporous layer. Pieces of this PTL were cut into rectangles of 1.8 x 3.8 cm$^2$, to match the single cell size used for DBFC measurements. Prior depositing the catalysts on the PTL, the pieces were cleaned in a piranha solution (mixture of concentrated sulfuric acid $H_2SO_4$ (98%) and hydrogen peroxide $H_2O_2$ (30%)). The PTL pieces, once cut, were immersed in the solution during 20 to 30 minutes, then rinsed with distilled water and dried with a stream of air. Besides eliminating all organic residues that could have been present on the surface or in the volume of the PTL, this treatment also improves the hydrophilic properties of the surface, allowing appropriate bounding of the catalysts on the substrate thanks to the presence of numerous hydroxyl groups. The piranha solution being an extremely dangerous solution, its preparation and its use were made under a fume hood in a chemistry room, while respecting all the safety principles required for such a solution.

**➔ Description of the synthesis process and preparation of the electrocatalysts**

The process for the synthesis of metallic nanoparticles and deposition on a substrate (Figure 1) is essentially the same for the three types of metals tested (Au, Pd and Pt); the only difference is that triangle-shaped particles are obtained for gold and sphere-shaped for palladium and platinum, with a higher density in the latter cases.

**- Gold nanotriangle electrodes (GNTE):** The process of synthesis of gold nanotriangles consists of dissolving a homopolymer, here PMMA, in a solution with PGMEA as a solvent, then adding gold salts ($HAuCl_4$) to this solution. PMMA acts as a matrix and allows uniform organization of gold nanoparticles on the surface of the substrate. Depending on the mass concentration of the PMMA solution, a high density of nanotriangles and some other minorly-encountered forms (nanosphere, nanopentagon, nanohexagon) of gold particles are observed on the substrate surface. The solution was sampled using a pipette and deposited on the carbon substrate after cleaning by spin-coating; then the samples were annealed at temperatures ≥ 200°C during 1 to 2 hours. After annealing, the surface of the samples was observed (by SEM) (Figure 2).

**- Palladium Nanoparticle Electrodes (PdNE):** The preparation of Pd nanoparticle electrodes uses the same solvent and polymer as for gold. The salts of palladium (II) nitrate hydrate ($PdNO_2$) were dissolved in a solution of PGMEA containing PMMA, the latter still acting as a matrix. In this process, whatever the mass concentration of the PMMA solution, the particles are of spherical shape. The solution was also sampled and deposited on the carbon as described above, then the substrate was annealed at temperatures ≥ 300°C during 1 to 2 hours. SEM images obtained for these PdNE are represented in Figure 3a and 3b.

**- Platinum Nanoparticle Electrodes (PtNE):** The preparation of Pt nanoparticle electrodes uses the same solvent and polymer as those for gold. In this case, the salts of platinum chloride ($PtCl_4$) were dissolved in a solution of PGMEA containing PMMA, which acts as a matrix. Whatever the mass concentration of the PMMA solution, spherical-shaped particles are obtained. The solution was sampled and deposited on the carbon substrate as for PdNE, the composite annealed at temperatures ≥ 300°C during 1 to 2 hours, and representative SEM images of the obtained materials were acquired (Figure 3c and 3d).

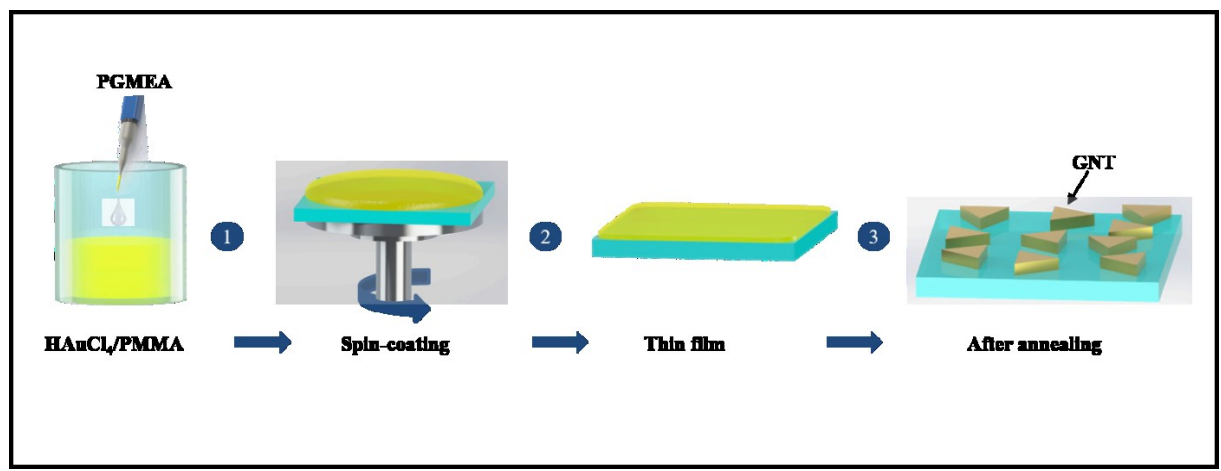

**Figure 1**: Fabrication process of GNT/C

### 2.2) Structural characterization

The synthesized electrodes were characterized using a high resolution FEG scanning electron microscope (SEM) (Hitachi SU8030 Tokyo, Japan) with an acceleration voltage between 5 and 10 keV and a working distance ≤ 10 mm. Crystallographic characterization was performed using X-ray Diffraction (XRD) using a 4-circles Bruker D8 Discover diffractometer equipped with Cu-Kα source (wavelength 0.154184 nm) and a 2D detector Pilatus-100K from Dectris. The crystallographic structure of the Pt/C electrocatalyst was further characterized by X-ray diffraction (XRD). Measurements were carried out on Bruker D8 Discover equipped with CuKα radiation (wavelength: 0.15418 nm, 40 kV, 40 mA), polycapillary optics and a collimator of 1 mm diameter. The detector was a Pilatus 100K (pixel size 172 μm, 195 x 487 pixels) located at 180.5 mm from the specimen. Acquisitions were performed from 35° to 90° in 2θ with a step size of 5° between two exposures and an acquisition time of 20s/exposure.

### 2.3) Electrochemical measurements

The tests at the level of the single cell were carried out using the experimental bench described in **[26, 27, 43]**. A pump was used to circulate the sodium borohydride solution with a concentration of 0.5 mol/L in 4 mol/L sodium hydroxide in the anode compartment of a cell with an active surface area of 6.8 cm$^2$ (Figure 4). The flow-field consists of a single serpentine channel, 1 mm wide and 0.7 mm deep. The cell is equipped with a hydrogen-fed reference electrode to measure the anode and cathode potentials. The cathode consists of a gas diffusion electrode (GDE, 0.3 mg/cm$^2$ Pt/C on a carbon cloth - fuelcellstore) hot-pressed on a Nafion 212 membrane and is supplied with pure oxygen. The home-made anode described above, using a is GDL SGL 28AA, is simply pressed against the membrane. Teflon flat gaskets allow the sealing and the GDE compression (by 20% of their thickness). Special care is taken to thermally control the cell and the borohydride solution, because the DBFC performance is very sensitive to the temperature. At the anode outlet, the mixture of the liquid solution and the hydrogen produced by hydrolysis are separated in a phase-separator. A flow of nitrogen is added and takes the hydrogen to a fuel cell operating as a hydrogen pump. By measuring the current, the flow of hydrogen present at the outlet can be deduced.

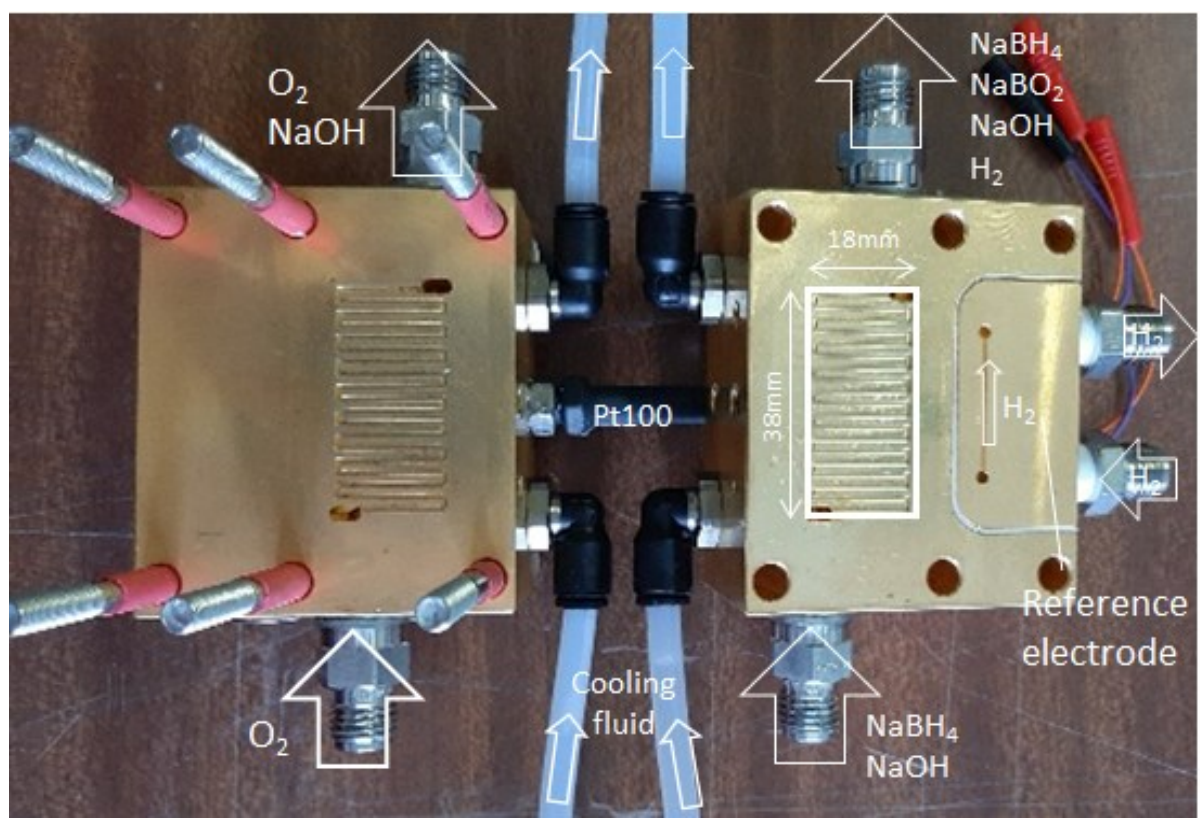

**Figure 4**: cell used to make the measurements. The temperature regulation is carried out by a water circulation controlled by a Pt100 probe. The borohydride solution is preheated before being injected.

### 3) RESULTS AND DISCUSSION

#### 3.1) Physical characterization

Chemical deposition has been used as a synthesis technique to prepare electrocatalysts for heterogeneous catalysis applications, in particular for the anodic reaction (BOR) in DBFCs. After annealing, the morphology of the PTL-supported catalysts was observed using (SEM) (Figure 2 and 3). The gold nanotriangle catalysts were prepared and deposited on the surface of the fibers and in the volume of the carbon substrate. Figure 2a and 2b correspond to the surface and the volume of the carbon fiber substrate without deposition of catalysts, respectively. The diameter of the fibers is approximately 6.5 µm. Figure 2c and 2d correspond to gold nanotriangles deposited on carbon fibers at different magnifications. The size of the nanotriangles is between 30 and 40 nm on average, obtained on 110 particles using the imageJ and OriginPro 2018 softwares. It is obvious that the nanotriangles and some other minorly-encountered forms (nanosphere, nanopentagon, nanohexagon, etc.) disperse uniformly at the fiber surface of the carbon support without any aggregation. The gold nanotriangles and the other shapes obtained on the surface are flat. Measurements using the atomic force microscope (AFM) were performed. The existence of other forms of particles is due to the presence of halide ions [I$^-$] which are incompatible with other ions [Au$^+$] in the solution **[40, 41]**. Another parameter which allows to improve the density of nanotriangles on the surface and in volume of the carbon substrate would be to perform successive deposits. By doing so, the particle density increases: Figure 2c and 2d correspond to 4 and 3 deposits respectively. An optimal number of deposits is required to limit the growth of the particles by the phenomenon of coalescence, though.

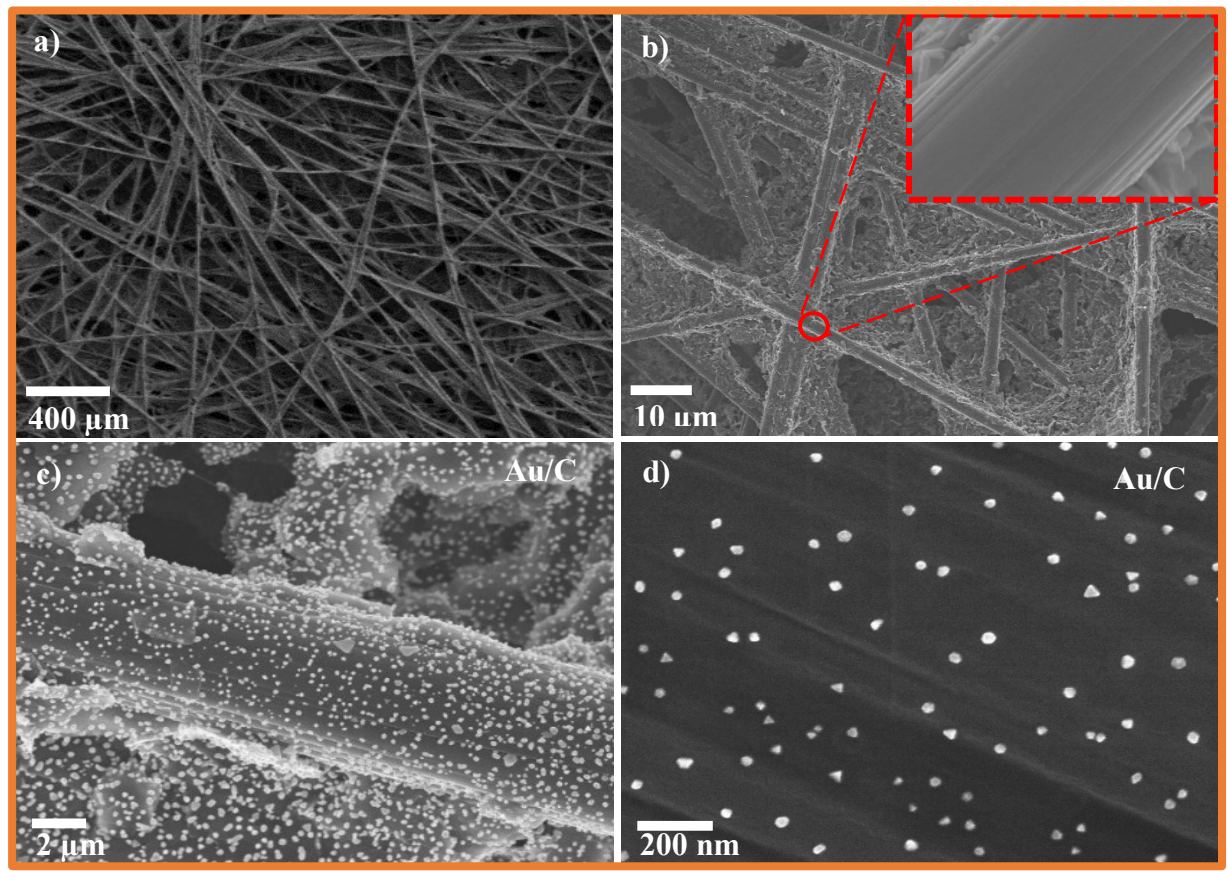

**Figure 2**: SEM images at different magnification. a) correspond to the surface of carbon fibers at x10k without deposit; b) corresponds inside the volume of the carbon support at x300k without deposit; c) and d) correspond to GNT deposited on surface of the carbon fibers at x50k and x400k respectively.

Similar to the preparation of Au/C electrocatalysts, Pd/C and Pt/C electrocatalysts were prepared by depositing Pd and Pt nanoparticles on the surface and in the volume of the carbon substrate. After annealing, the morphology of the PTL-supported catalysts was imaged by SEM (Figure 3). The Pd and Pt nanoparticles exhibit a spherical morphology on the surface of the carbon substrate fibers with uniform dispersion. Figures 3a and 3b correspond to the Pd nanoparticles deposited on the surface of the fibers of the carbon substrate at different magnification, while Figures 3c and 3d correspond to those of Pt. The average size of the Pd particles is between 2 and 7 nm obtained on 246 particles; that of Pt is near identical to that of Pd, between 2 and 8 nm obtained on 92 particles. Some areas of aggregation of somewhat larger particles can occasionally be seen in Figures 3a and 3c, due to the rough surface of the carbon fibers (but agglomerates are not predominant). Overall, the technique enables producing isolated nanoparticles of controlled size, well-dispersed in the whole volume and on the whole surface of the carbon PTL. From now-on, emphasis is made on the Pt/C sample for further characterizations.

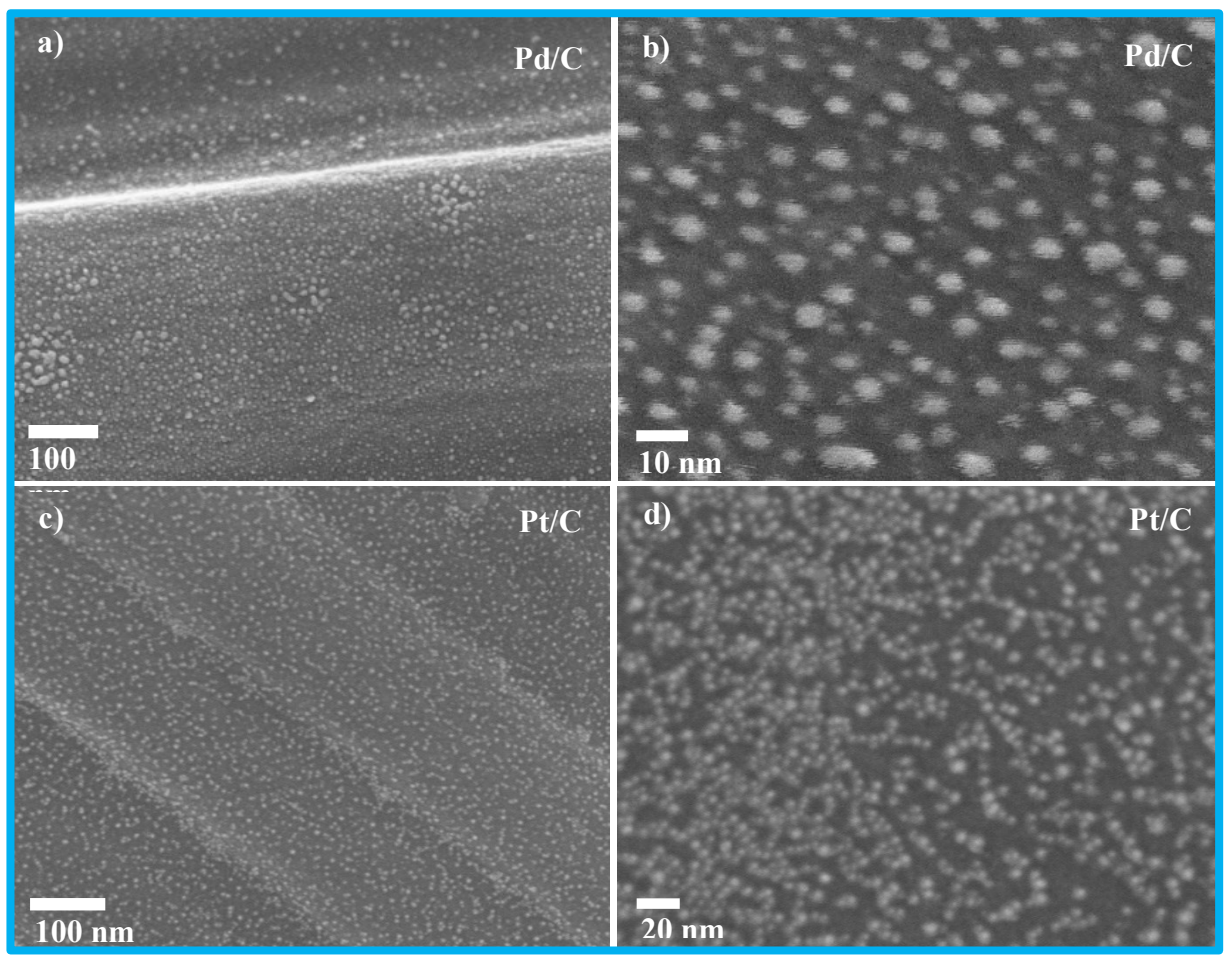

**Figure 3**: SEM images of NPs Pd and Pt on porous transport layer (PTL). a) and b) correspond to Pd NPs deposit on surface of the carbon fibers at x100k and 800k respectively; c) and d) correspond to Pt NPs deposited on surface of the carbon fibers at x110k and x300k respectively.

The experimental XRD pattern of the elaborated electrode is shown in Figure 5 (in magenta). On this diagram the diffraction peaks of face centered cubic (fcc) platinum (blue dotted lines, JCPDS file No. 04-0802) and graphite (red dotted lines, JCPDS file No. 75-1621) can be observed. The peaks from Pt are indexed on Figure 5; they correspond to the lattice planes (111), (200), (220) and (311), respectively. The other diffraction peaks correspond to the carbon graphite substrate.

The crystallite size of platinum was determined from the diffraction peak width $\beta$ using Scherrer's equation **[42]**. The peak width was obtained through a least squares fitting with a pseudo-Voigt function, after removal of the background and Lorentz-Polarization correction. The instrumental broadening was estimated with the help of NIST 1976B corundum reference standard, measured with the same instrumental set-up, and was removed from the platinum peaks. The crystallite size obtained, averaged on the four peaks is 9.9 nm with a standard deviation of 0.7 nm. This analysis neglects the broadening due to the imperfections of the crystal lattice (lattice distortion, also called rms strain). Thus, the value of 9.9 nm is slightly overestimated. However, it was found that the size estimated on the first 3 peaks is almost constant which means that the lattice distortion remains quite low, and the average crystallite size agrees well with the NPs sizes measured by SEM.

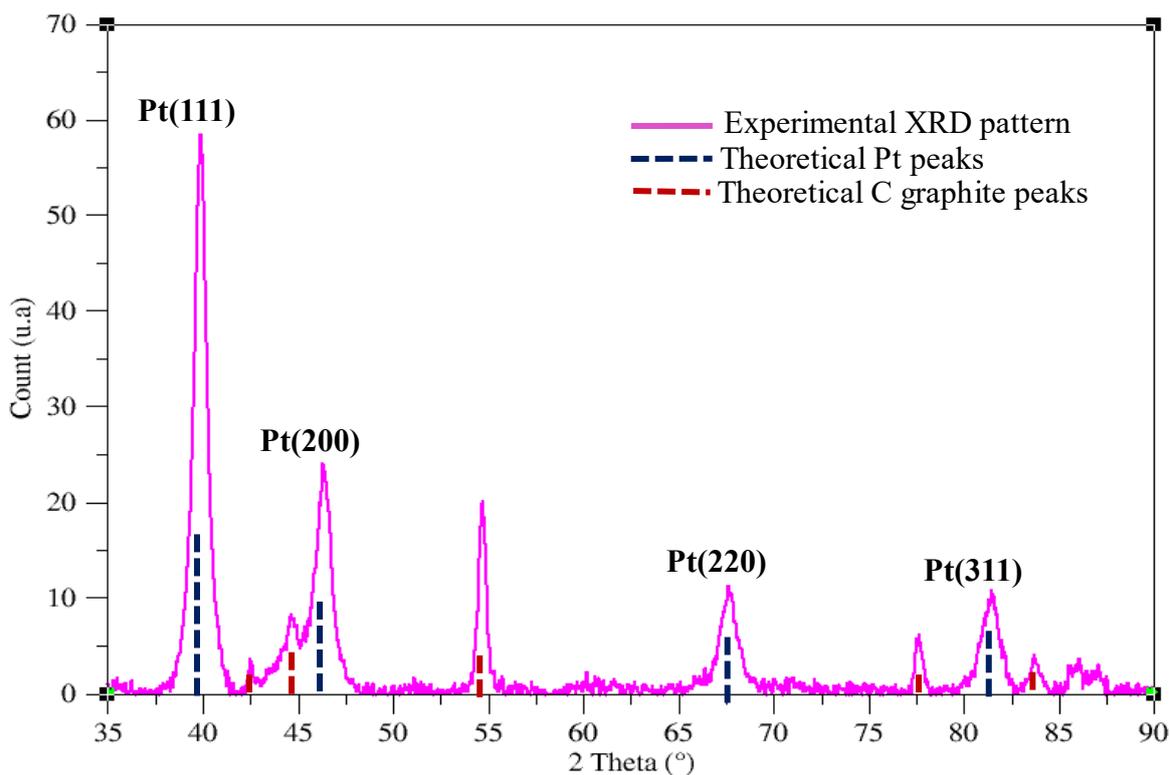

**Figure 5**: XRD pattern of Pt NPs on carbon fibers obtained with a Bruker ® D8 Discover diffractometer, compared to the theoretical patterns (blue lines for Pt and red lines for carbon fibers.

### 3.3) Fuel cell performance

The results obtained for a temperature of 60°C, three platinum loadings (0.16, 0.32 and 0.48 mgPt/cm$^2$) and the nickel electrode described in **[27]** are shown in Figure 6. Relatively similar performances are observed whatever the platinum loading and even the nature of the electrode. A maximum power density slightly lower than 200 mW/cm$^2$ is observed at a current density of 0.4 A/cm$^2$. These similar performances are due to the fact that with these optimized electrodes, the limiting factor of the cell performance is the membrane resistance due to the transport of sodium Na$^+$ ions: 0.8 Ω.cm$^2$ at 60°C, this resistance being too high to expect a power density exceeding 200 mW/cm$^2$. The cathode can also be limiting, due to the production of NaOH, which reduces the mass-transport of oxygen to the catalytic sites **[43]**. This means that the platinum loading at the anode can certainly be further reduced without compromising the DBFC performance. It can also be seen on Figure 6 (right panel), that the higher the platinum loading, the greater the hydrogen escape. It is therefore understandable why it is useful to manufacture electrodes with a sufficiently low loading, to improve the faradic efficiency, the loading being ideally high enough to not compromise the power output. For a loading of 0.16 mgPt/cm$^2$, the hydrogen escape becomes negligible above a current density of 0.3A/cm$^2$. In that sense, the low-loading electrodes prepared herein, enable to reach larger faradic efficiency than nickel electrode prepared in **[26, 27]**, which are limited to about 50% faradic efficiency whatever the current density. This shows the benefit of the present elaboration strategy of metallic nanoparticles for heterogeneous catalysis/electrocatalysis applications.

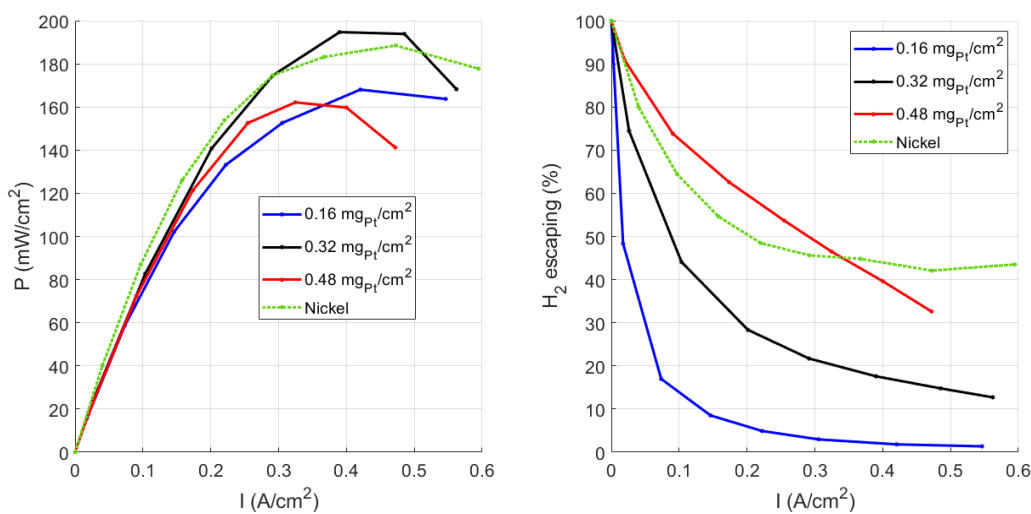

**Figure 6:** Left, electrical power output for three platinum charges and the nickel electrode described in **[27]**. Right, corresponding output hydrogen flow rates as a function of the current density (so-called hydrogen escape, expressed in % of the generated current density).

### 4) CONCLUSION

In summary, a single step chemical synthesis process has been developed for the fabrication and direct deposition of catalysts on the surface and in the volume of porous carbon substrates. The synthesis process simply consists of dissolving in the same solvent (PGMEA) a metal precursor ($HAuCl_4$, or $PdNO_2$ or $PtCl_4$) containing a homopolymer (PMMA), then depositing this solution on the porous carbon fiber substrate. With this methodology, gold nanotriangles (ranging from 30 to 40 nm in size) are obtained depending of the mass concentration of the PMMA solution, while nanospheres are obtained for Pd and Pt ranging from 2 to 7 nm and 2 to 8 nm in size, respectively. The catalysts are uniformly dispersed on the surface and in the volume of the carbon support with a very low extent of aggregation, and present homogeneous size and shape distribution. Platinum-based electrodes were prepared for the anodic reaction in a DBFC. Three platinum loadings (0.16; 0.32; and 0.48 mgPt/cm$^2$) were studied and benchmarked against a state-of-the-art nickel electrode. These platinum-based electrodes demonstrated improved Faradic efficiency, especially with a low platinum loading (0.16 mgPt/cm$^2$) compared to the nickel electrode. An optimum platinum loading was noticed, that is the best compromise between power density and faradic efficiency. These results, demonstrated for the example of DBFC anodes, illustrate that this very simple synthetic process shows promise for the preparation of numerous types of metal catalysts on various supports, which may have potential applications in heterogeneous catalysis and electrocatalysis.

**ACKNOWLEDGMENT:**

This work has been supported by SAYENS (NATO project) and by the Agence Nationale de la Recherche and the FEDER (INSOMNIA project, contract "ANR-18-CE09-0003 »). Financial support of NanoMat (www.nanomat.eu) by the "Ministère de l'enseignement supérieur et de la recherche," the "Conseil régional Champagne-Ardenne," the "Fonds Européen de Développement Régional (FEDER) fund," and the "Conseil général de l'Aube" is also acknowledged.